\documentclass{aastex63}
\shorttitle{NS phase transition as the origin of FRBs and SGRs}
\shortauthors{Shen et al.}
\graphicspath{{./}{figures/}}
\usepackage{booktabs}
\usepackage{amsmath}
\usepackage{subfigure}
\usepackage{cancel}
\begin{document}

\title{Neutron star phase transition as the origin for the fast radio bursts and soft gamma-ray repeaters of SGR J1935+2154}

il to set provide email addresses. Each \email will appear on its

\correspondingauthor{Yuan-Chuan Zou}
\email{zouyc@hust.edu.cn}

\author[0000-0002-7949-3906]{Jun-Yi Shen}
\affiliation{School of Physics, Huazhong University of Science and Technology, Wuhan 430074, China}


\author[0000-0002-5400-3261]{Yuan-Chuan Zou}
\affiliation{School of Physics, Huazhong University of Science and Technology, Wuhan 430074, China}

\author[0000-0002-5319-8386]{Shu-Hua Yang}
\affiliation{Institute of Astrophysics, Central China Normal University, Wuhan 430079, China}

\author{Xiao-Ping Zheng}
\affiliation{Institute of Astrophysics, Central China Normal University, Wuhan 430079, China}

\author[0000-0003-4976-4098]{Kai Wang}
\affiliation{School of Physics, Huazhong University of Science and Technology, Wuhan 430074, China}



\begin{abstract}
Magnetars are believed as neutron stars (NSs) with strong magnetic fields. X-ray flares and fast radio bursts (FRBs) have been observed from the magnetar (soft gamma-ray repeater, SGR J1935+2154). We propose that the phase transition of the NS can power the FRBs and SGRs.
Based on the equation of state provided by the MIT bag model and the mean field approximation, we solve the Tolman-Oppenheimer-Volkoff equations to get the NS structure. With spin-down of the NS, the hadronic shell gradually transfers to the quark shell.
The gravitational potential energy released by one time of the phase transition can be achieved. The released energy, time interval between two successive phase transitions, and glitch are all consistent with the observations of the FRBs and the X-ray flares from SGR J1935+2154. We conclude that the phase transition of an NS is a plausible mechanism to power the SGRs as well as the repeating FRBs. 
\end{abstract}


\section{Introduction} \label{sec:intro}
Since the discovery of fast radio burst (FRB) in 2007 \citep{2007Sci...318..777L}, the origin and the radiation mechanism are still mysterious \citep[see][for recent reviews]{2019A&ARv..27....4P,2019PhR...821....1P,2020Natur.587...54C}.
While for the non-repeating FRBs, they are still possible from catastrophic events, such as collapses of supra-massive neutron stars (NSs) to black holes (BHs) \citep{2014A&A...562A.137F,2014ApJ...780L..21Z}, mergers of charged BHs \citep{2016ApJ...827L..31Z}, mergers of binary white dwarfs (WDs) \citep{2013ApJ...776L..39K}, WD-NS binaries \citep{2016ApJ...823L..28G}, NS–NS mergers \citep{2018PASJ...70...39Y}, NS-astroid mergers \citep{2016ApJ...829...27D}, dark matter-induced collapses of NSs \citep{2015MNRAS.450L..71F}, the evaporation of primordial BHs \citep{2012MNRAS.425L..71K} and so on. The discovery of the repeating FRB 20121102 \citep{2016Natur.531..202S} suggests that at least the repeating FRBs are very likely coming from a certain kind of energy release from NSs \citep[see][for example]{2020MNRAS.498.1397L}. Especially with the discovery of periodic repeating FRBs, such as FRB 20180916 \citep{2020Natur.582..351C}, FRB 20121102 \citep{2020MNRAS.495.3551R}, various models have been proposed for the periodicity \citep{2020ApJ...893L..31Y, 2020ApJ...893L..26I, 2020ApJ...892L..15Z, 2020MNRAS.496.3390B, 2020PASJ...72L...8C, 2020MNRAS.497.1001S, 2020RAA....20..142T, 2021ApJ...917...13S, 2021ApJ...918L...5L, 2021ApJ...922...98D, 2022A&A...658A.163W}, while all of them are more or less related to the NSs.

A special FRB was discovered in 2020, FRB 20200428 in the Milky Way galaxy \citep{2020Natur.587...54C}. It was found associated with a soft gamma-ray repeater (SGR), SGR J1935+2154 (it is also called SGR 1935+2154 in some literature), which was first discovered by \citet{2016MNRAS.457.3448I}.
Recently, the Canadian Hydrogen Intensity Mapping Experiment (CHIME)/FRB Collaboration reported that they detected another radio burst from the direction of SGR J1935+2154 in its active window \citep{2022ATel15681....1D}. It is also reported reactive in X-ray band \citep{2022ApJS..260...24C}. These indicate that the FRBs may connect with SGRs, and both of them are repeated. The energy source of both phenomena may come from the NS.

The structure of an NS can be divided into five regions, the atmosphere, the envelope, the crust, the outer core, and the inner core. The density of the crust can reach  $\rho \sim 10^{11}\mathrm{g/cm^{3}}$. In this region, partial neutrons do not form nuclei and they become unbound from nuclei \citep{1971ApJ...170..299B}. The perturbative quantum chromodynamics (QCD) shows that dense matter $\rho \sim 40\rho_s$  consists of asymptotically free quark matter, where $\rho_s\approx 2.7\times10^{14}\mathrm{g/cm^{3}}$ \citep{PhysRevD.81.105021}. The highest density in an NS's inner core is about 10$\rho_s$ \citep{2007PhR...442..109L}, but some observations suggest that there is the stiff equation of state (EOS) \citep{BURGIO2021103879} in the NSs' core. This means that quark deconfinement could appear in NSs.
\citet{2022arXiv221104662U} found the phase transition can go back and forth in the numerical simulation. \citet{2018ApJ...858...88Z} even suggested that the repeating FRB can be powered by the collapsing of hadronic matter accreted onto a strange star crust.

Collapsing of the crust may induce glitches of the spin. Glitch is a common phenomenon in pulsars, e.g., the glitch detected in the anomalous X-ray pulsar 1RXS J170849.0-400910 \citep{2000ApJ...537L..31K}. 
The spin frequency $\nu$ increases roughly in the range of $\frac{\delta \nu}{\nu} \approx 10^{-5} \sim 10^{-12} $. \citet{2011RAA....11..679X} suggested that the glitch could be caused by the phase transition of hadronic matter in the NS core. The inner pressure of the NS can increase due to some reasons, such as the spin-down of the NS. The hadronic matter may transit into quarks and the star's structure gets changed. This can generate an accompanying starquake, and consequently, the NS's moment of inertia is reduced, inducing a glitch due to the conversation of spin angular momentum. After this process, the gravitational potential energy will be released. This energy could be the source of FRBs and X-ray bursts (XRBs) in a certain active period of an SGR. \citet{2018ApJ...852..140W} discussed that starquakes could produce FRBs, and argued that the energy released from the gravitational potential energy released during a type-II starquake can even reach $\sim 10^{47}\mathrm{erg}$. The released energy is quite adequate for one FRB event. 

In this work, we discuss the phase transition as the energy source for the FRBs and X-ray bursts (XRBs) from an SGR, based on the observations of the SGR J1935+2154. The paper is arranged as follows. In Section 2, we show how to get the EOS of hadronic matter and quark matter, eventually achieving the NS's structure. In Section 3, we show the phase transition condition. In Section 4, we calculate the phase transition and apply it to SGR J1935+2154. The conclusion and discussion are given at the end.

\section{Hadronic and Quark phase }\label{2}
In high-energy particle collision experiments, people have predicted that protons and neutrons, which are the composition of three quarks ({\it uud} and {\it udd}, where {\it u} denotes up quark, and {\it d} denotes down quark), in the high density and the high temperature will be broken and form quark-gluon plasma (QGP) \citep{2018ARNPS..68..339B}. The NS core is very dense, which is naturally a high-energy physics laboratory. Some observations on the massive pulsar PSR J0740+6620 \citep{2021ApJ...918L..27R} suggest that the EOS has a stiff component in 2-4$n_s$ \citep{2018ARNPS..68..339B} ($n_s$ is about $0.163 \,\mathrm{fm^{-3}}$, the nucleon saturation density), which could be the interface phase EOS. This strongly suggests that the inner NS has the appearance of deconfined quarks.
\par
\subsection{Hadronic phase}
There are many different models that describe the hadron and quark phases \citep{2016ARA&A..54..401O}, and each model corresponds to a different EOS.  The EOS is the relation of pressure $p$ and density $\rho$. Because the NS structure is related to the EOS, phase transition will induce the change of the star's EOS, variation of the structure, and glitches.  \citet{2022MNRAS.516.1127P} 
 studied the spin-down effect, which could be a trigger for phase transition. \citet{2011RAA....11..679X} studied the phase transition process. We chose a similar way to investigate the phase transition. 
 
We choose several model of hadronic matter, such as DD2 \citep{PhysRevC.81.015803,PhysRevC.90.055203}, DS(CMF)-5 \citep{2008ApJ...683..943D,PhysRevC.92.055803}, VQCD(APR) \citep{2019JHEP...07..003I,PhysRevC.58.1804} to describe the hadronic matter and the crust. The DD2 model is based on the relativistic mean field (RMF) theory, which explains the strong interaction between nucleons by exchanging mesons.  Replacing the operators by their spacetime, constants, mean values, and nuclei by ground state expectation generated in mean meson fields, and the fields are treated as classic fields. The Lagrangian density is first
 applied, followed by deriving the equation of motion and combining the assumptions to obtain the energy density and pressure. more details see \citep{PhysRevC.90.055203}.

\subsection{Quark phase}
\begin{figure}
\subfigure[]{
\label{Fig.sub.1}
\includegraphics[height=0.38\linewidth]{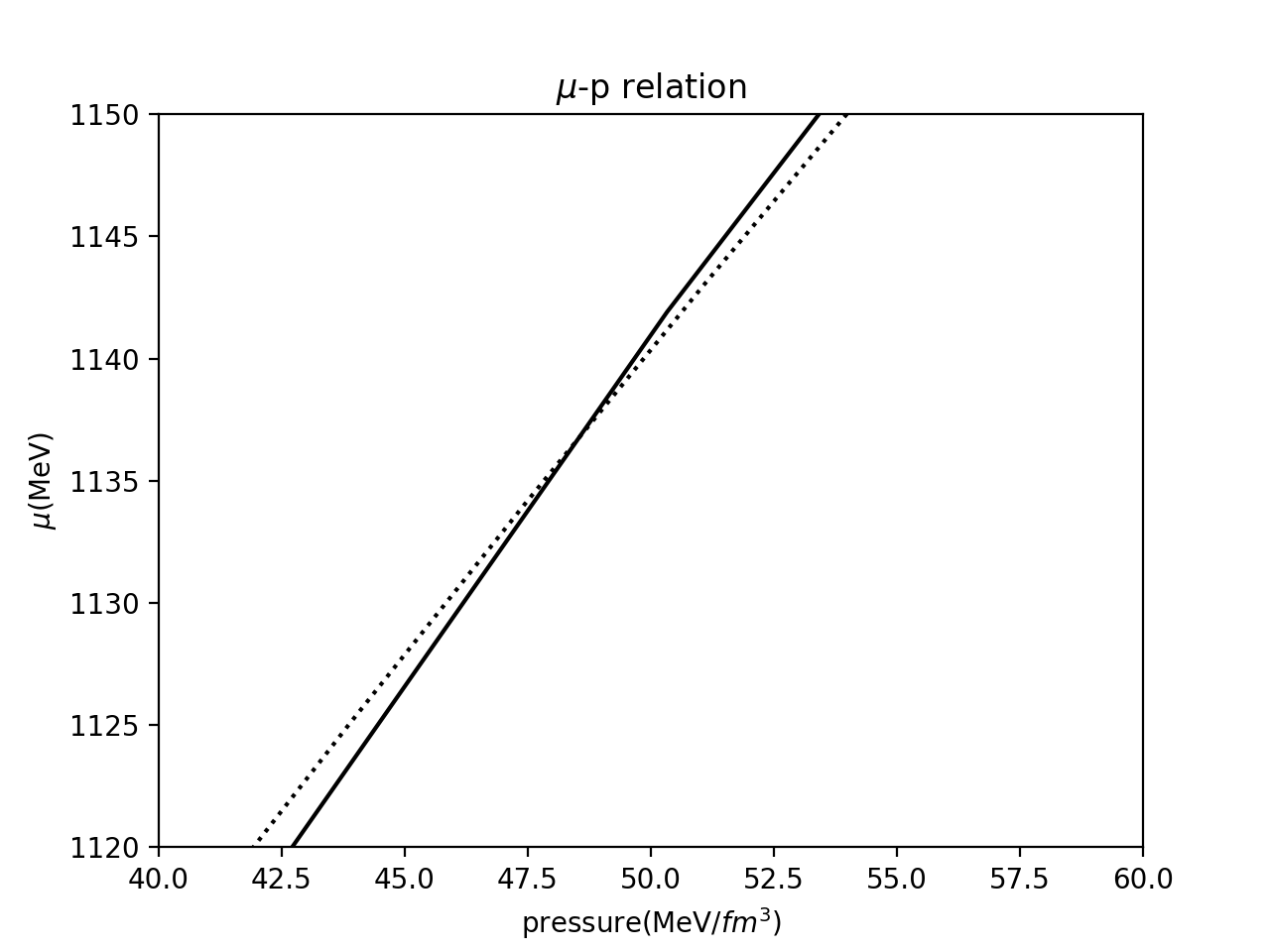}}
\subfigure[]{
\label{Fig.sub.2}
\includegraphics[width=0.45\linewidth]{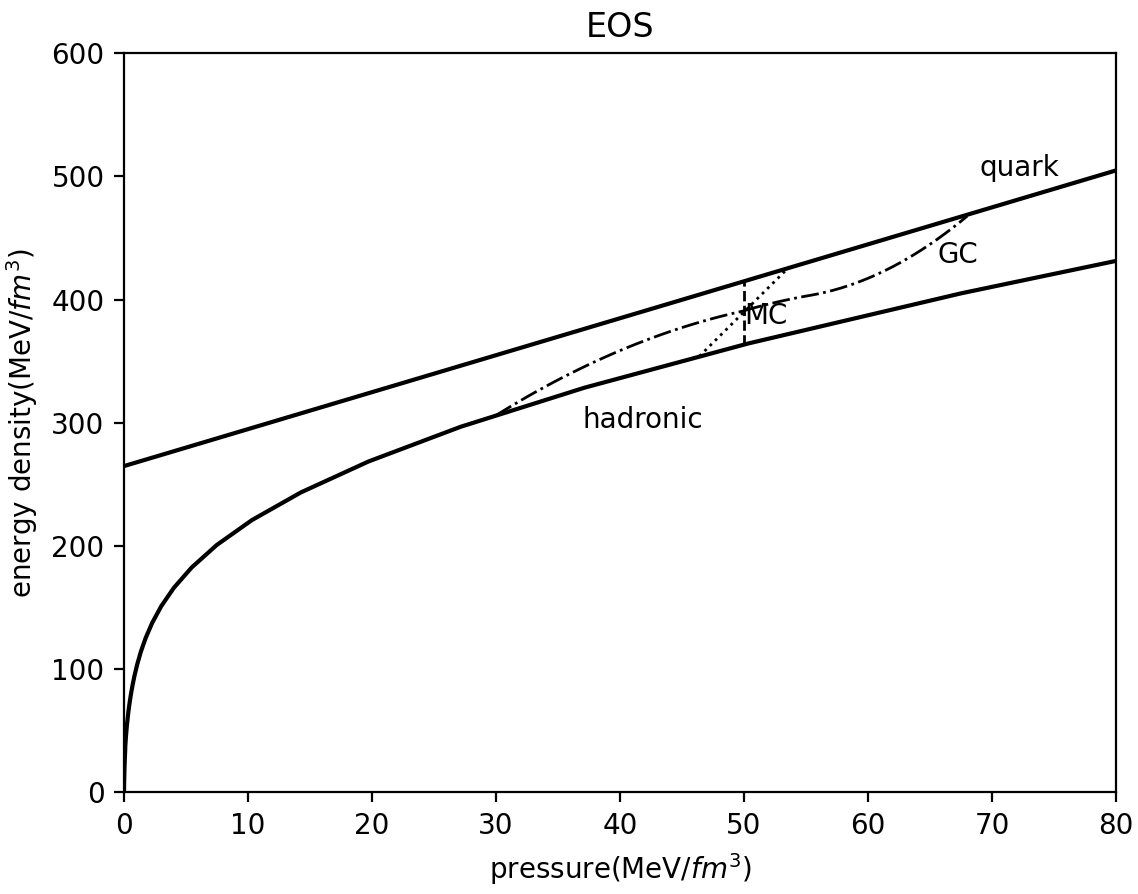}}
\caption{(a). Chemical potential $\mu$ $-$ pressure $p$ diagram for hadronic matter described by DD2 model (black solid line) and quark matter described by MIT bag model (dotted line), which are denoted in the figure. The two lines were calculated based on the public data  in  \href{https://compose.obspm.fr}{https://compose.obspm.fr}, and more details are shown in \citet{2015PPN....46..633T}. They cross each other at $p_c \simeq  \boldsymbol{48.53} \mathrm{MeV/fm^{3}}$. As discussed in Section \ref{3}, the matter will stay in a low chemical state, therefore, the matter will suffer a phase transition at around $ \boldsymbol{48.53} \mathrm{MeV/fm^{3}}$, with pressure changing. This phase transition is called MC transition. 
(b). EOS diagram for different conditions, which are denoted in the figure. The data used here is also from \href{https://compose.obspm.fr}{https://compose.obspm.fr}. Three components of EOS for NSs, hadronic matter calculated by DD2, and quark matter with the MIT bag model. The two solid lines correspond to quark matter and hadronic matter respectively. The dashed vertical line denotes the MC model, which corresponds to the cross point in panel (a). The dashed-dotted line is for the GC model. We did not calculate the GC curve, and GC shown here is schematic. The dotted line is for the mixed model with a free parameter $k$, where $k$ is the slope of the line.
\label{fig1}}
\end{figure}

Quarks are simply described by the MIT bag model . With this model, the quark matter pressure and energy density are \textbf{ \citep{2005ApJ...629..969A}}:
\begin{equation}
\begin{split}
 &  p = -C +  \frac{1}{4 \pi^{2}} a_4 \mu^4 ,\\
 &   \epsilon   =  4C + 3p .
\end{split} \label{eq:quarkPhase}
\end{equation}
We take  $C^{\frac{1}{4}}=150$MeV, \textbf{and $\boldsymbol{a_4=0.56}$}. For any total baryon number density, the two reaction equilibria are 
\begin{equation}  \label{eq:mup}
\begin{split}
    &\mu_n = \mu_p + \mu_e, \\
    &\mu_d = \mu_u + \mu_e.
\end{split}
\end{equation}
\textbf{and the quark chemical potential $\mu$} is

\begin{equation}  \label{eq:mupp}
\begin{split}
    &  \boldsymbol{ \mu \equiv \mu_d- \frac{1}{3} \mu_e = \frac{1}{3}(\mu_u+2\mu_d) .}\\
\end{split}
\end{equation}
With these setups, we get the quark matter EOS, as shown in Fig \ref{Fig.sub.2}. The quark's mass is far less than its Fermi momentum. Thus, the EOS is proximate linear. EOS of matter goes the quark line in Fig \ref{Fig.sub.2} only if the pressure is larger than  $ {\mathrm{\boldsymbol{48.53}MeV/fm^{3}}}$.

\section{Phase transition}\label{3}

\subsection{Reaction equilibrium}
Assuming $\alpha$ and $\beta$ denote two phases, they consist of an isolated system. The second law of thermodynamics tells us when the system gets balanced, $\delta S=\delta S _{\alpha}+\delta S _{\beta}=0$. This equals to $\delta U_{\alpha}(\frac{1}{T_{\alpha}}-\frac{1}{T_{\beta}})$+$\delta V_{\alpha}(\frac{p_\alpha}{T_{\alpha}}-\frac{p_\beta}{T_{\beta}})$-$\delta n_{\alpha}(\frac{\mu_\alpha}{T_{\alpha}}-\frac{\mu_\beta}{T_{\beta}})=0$. Thus, the three balance conditions are mechanical equilibrium $p_{\alpha}=p_{\beta}$, thermo-equilibrium $T_{\alpha}=T_{\beta}$, and chemical equilibrium $\mu_{\alpha}=\mu_{\beta}$, respectively. 
For the evolving condition, this $\delta S$ should be larger than zero. To guarantee the entropy of the isolated system increases, the evolution must obey $\delta n_{\alpha}(\frac{\mu_\alpha}{T_{\alpha}}-\frac{\mu_\beta}{T_{\beta}})<0$ if the temperature and pressure of two phases are equal. This inequality means if the chemical potential of phase $\alpha$ is larger than the chemical potential of phase $\beta$, $\delta n_{\alpha}<0$, the particle number in phase $\alpha$ will decrease and particles in phase $\alpha$ transit into phase $\beta$. Which phase appears depends on the corresponding chemical potential. Matter prefers to stay in the lower chemical potential state and this is the lower energy state. 
In Fig. \ref{Fig.sub.1}, we show the $\mu-p$ relation of the two phases. The cross point that satisfies  $\mu_{\mathrm{H}}(p_c,T)=\mu_{\mathrm{Q}}(p_c,T)$, $p_c= \boldsymbol {48.53} \mathrm{MeV/fm^{3}}$. 
For $p<p_c$, as $\mu_{\mathrm{H}}(p_c,T)<\mu_{\mathrm{Q}}(p_c,T)$, matters will stay in the hadronic phase. For the case $p>p_c$, as $\mu_{\mathrm{H}}(p_c,T)>\mu_{\mathrm{Q}}(p_c,T)$, matters will stay in the quark phase. This phase transition condition is called the Maxwell construction (MC) phase transition \citep{PhysRevD.88.063001}.  However, this is an ideal condition. The real condition is more complex, for example, the metastable state because of the surface tension \citep{1983A&A...126..121S,1987A&A...172...95Z}. We will discuss the surface tension effect in the next subsection. 

\subsection{Metastable state}\label{hehehe}
We calculate the cross point $p_c$ in MC.
If the EOS of phase transition obeys MC, only infinitesimal  matter will transit. This tiny variation cannot be observed. There should be a metastable shell, where substantial particles take part in the phase transition. We should consider the interface phase. For a sphere interface, surface tension is $2\sigma/ r$, where $r$ is the sphere's radius.
The mechanical equilibrium should be rewritten as $p_{\mathrm{Q}} + 2 \frac{\sigma}{r} = p_{\mathrm{H}}$, where $p_{\mathrm{Q}} \equiv p'$, and the superscript H and Q denote hadron and quark. The chemical equilibrium formula also changes to $\mu_{\mathrm{H}} (p' + \frac{2 \sigma}{r},T) = \mu_{\mathrm{Q}} (p',T)$. The approximate solution is 
 $(p' - p_c + \frac{2 \sigma}{r}) V^\alpha_m = RT \ln \frac{p'}{p_c}$, where $V^\alpha_m$ is the mole volume. Then, we get the critical radius 
\begin{equation}
 r_{cr} = \frac{2 \sigma V^\alpha_m}{RT \ln \frac{p'}{p_c}}.
\end{equation}

In the region where the pressure is larger than $p_c$, quark matter appears in the hadronic matter. Here we assume the quark matter is the stable phase, i.e., it forms sphere droplets. If the quark droplet radius is larger than the critical radius $r_{c r}$, because $\mu_{\mathrm{H}}(p'+\frac{2\sigma}{r},T)<\mu_{\mathrm{H}}(p'+\frac{2\sigma}{r_{cr}},T)=\mu_{\mathrm{Q}}(p',T)$, the droplet will grow larger and larger. $r_{cr}$ should be small so that the phase transition is easy to proceed. Otherwise, the droplet will get smaller and disappear. Thus, even in $p>p_c$ region, the matter still can stay in the hadronic phase. This is the so-called metastable state. $p_c \rightarrow p'$ region is metastable. The quark droplets formed because of fluctuation, so the size  of the droplets  is random. The radius of quark droplets is unknown, but it is related to surface tension, and we can select a representative value of $p'$.

\subsection{EOS of NS}
Quarks may deconfine in high temperatures and high densities. There could be deconfined quarks at the inner region of the NS \citep{2018ARNPS..68..339B}. The closer to the core of the NS, the possibility of quarks appearing higher. For resisting gravitation, the pressure gradient is negative. With the inner radius, the pressure is higher. In the ideal condition (neglecting the surface tension), there is a critical radius $r_c$ in NSs. Neutrons stay in the hadronic phase at position $r>r_{c}$ and they stay in the quark phase at position $r<r_c$. 
The property of $r_c$ is that $\mu_\mathrm{H}(r_c)=\mu_\mathrm{Q}(r_c)$, which corresponds to the cross point in Fig. \ref{Fig.sub.1}. People   believed that the hadron-quark phase transition is a first-order phase transition \citep{1992NuPhA.540..630R}. Thus, in NSs, the quark matter and the hadronic matter have a very thin interface. However, \citet{PhysRevD.46.1274} took the mixed phase into consideration, and there could be a mixed phase in NSs. The mixed phase is constructed by the Gibbs construction (GC) \citep{PhysRevD.88.063001}. \citet{2000NuPhA.677..463S} discussed the mixed phase in detail. He ignored the Coulomb force and interface phase effect. Following \citet{2000NuPhA.677..463S}, we can get the EOS of mixed phase constructed by the GC, which is shown in Fig .\ref{Fig.sub.2}. The surface tension of the interface phase, $\sigma$ has an impact on the property of the mixed phase.  At present, the study of $\sigma$ is poor. People take out different models, giving large $\sigma$, $\sigma \approx 420 \mathrm{MeV/fm^{2}}$  \citep{2001PhRvD..64g4017A}, and small $\sigma$, $\sigma \approx60 \mathrm{MeV/fm^{2}}$ \citep{PhysRevC.99.035804}.  We take it as a free parameter in this article, same as in \citet{2011RAA....11..679X}. MC is the zero surface tension limit. GC is the infinite surface tension limit. MC corresponds to a really stiff EOS and GC relates to a soft EOS. The dotted line in Fig. \ref{Fig.sub.2} describes the mixed phase, and its slope $k$ is dependent on the surface tension $\sigma $. As the $\sigma$ is a free parameter, it leads $k$ as a free parameter. For a finite value of $\sigma$, the $k$ is in the range of [0,$+\infty$]. The EOS of mixed-phase is between GC and MC. Thus, the EOS of mixed-phase should cross the point MC and GC's intersection point. We did not calculate the GC, and we simply take three typical energy density values of cross point  $[3\epsilon_{\mathrm{H}}(p_c) + \epsilon_{\mathrm{Q}}(p_c) ] / 4, [ \epsilon_{\mathrm{H}}(p_c) + \epsilon_{\mathrm{Q} } (p_c)] / 2, [\epsilon_{\mathrm{H}}(p_c)+3\epsilon_{\mathrm{Q}}(p_c)]/4$, \textbf{ corresponding to $\epsilon_1$, $\epsilon_2$ and $\epsilon_3$ in Table \ref{tab}, respectively. This simplification makes sense since the shape of the Gibbs construction (GC) curve ensures that the cross point will not be located close to the two end points of the mixed phase line (see GC calculation shown in \citet{2011RAA....11..679X},  \citet{PhysRevC.99.065802} and \citet{constantinou2023framework}). Therefore, this simplification does not significantly affect our conclusions.} The EOS of NSs is divided into three parts, hadronic matter, mixed phase, and quark matter. In Fig. \ref{Fig.sub.2}, the mixed phase is described by slope $k$. Different $k$ will induce different amounts of matter in the metastable state.

\begin{figure}
    \centering
    \includegraphics[width=0.55\linewidth]{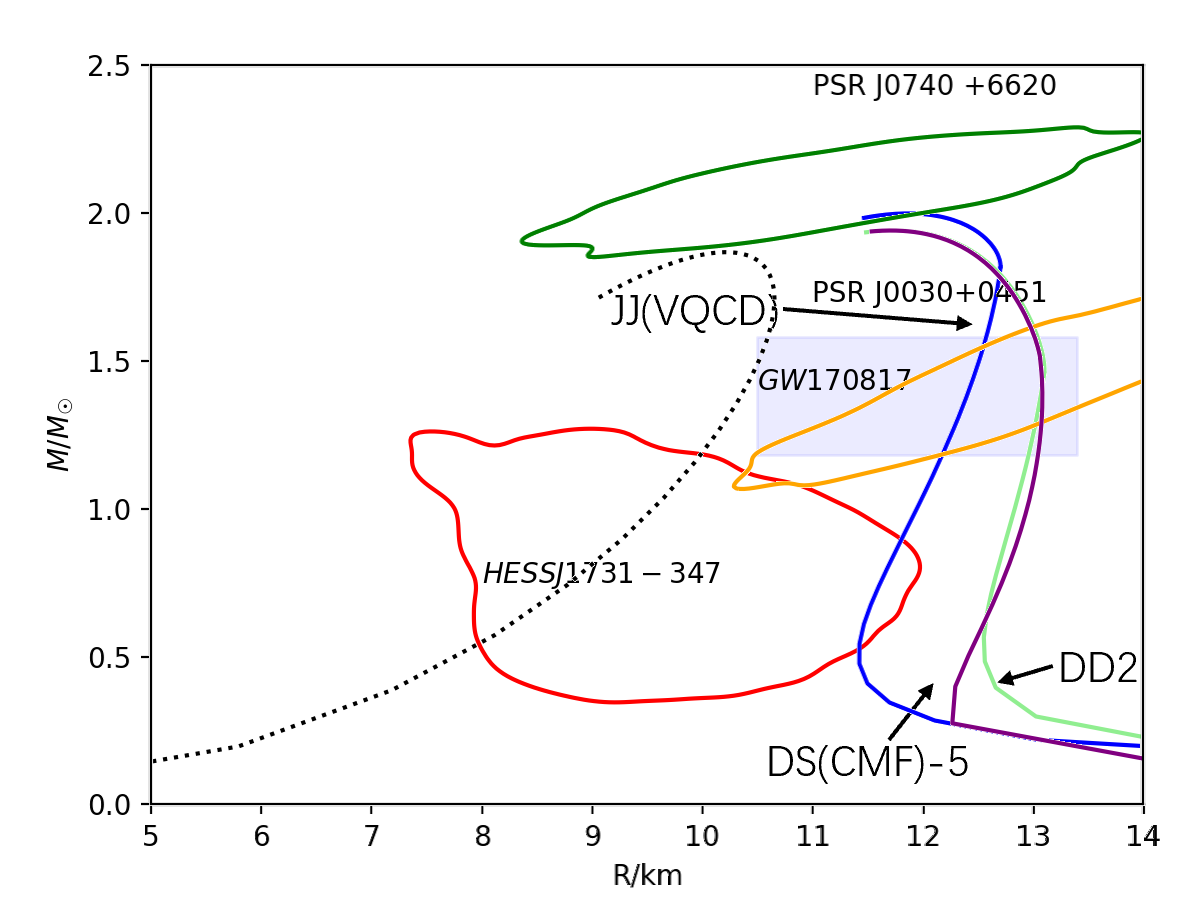}
    \caption{The black dotted line denotes pure quark M-R relation without spin, and the blue line (JJ(VQCD) model for hadronic matter), purple line (DS(CMF-5) model for hadronic matter), and the light green line (DD2 model for hadronic matter) denote hybrid star M-R relation without spin. The very large value of $k (> 10^{3})$ and energy density at cross point has no significant effect on M-R relation \textbf{because EOSs are almost the same. All EOSs are approaching the EOS of MC in this condition that $k (> 10^{3})$ is huge. } Thus, all the lines are for $k=2\times 10^9$ and density equals to $[\epsilon_{\mathrm{H}}(p_c) + \epsilon_{\mathrm{Q} } (p_c)] / 2$). \textbf{The 90\%  confidence regions are from the observation of HESS J1731-347 (enclosed by the red line) \citep{2022NatAs...6.1444D}, from the observation of PSR J0740+6620  (enclosed by the green line) \citep{2021ApJ...918L..27R}, and from the observation of PSR J0030+0451  (enclosed by the orange line) \citep{2019ApJ...887L..24M}, respectively.}  The light blue area represents GW170817 constraints from \citet{PhysRevLett.121.161101}. }
    \label{fig:my_label}
\end{figure}

There are many different EOSs describing neutron stars structure.  To ensure that the EOS we choose is viable, we need to consider the following astrophysical constraints.
\begin{itemize}
    \item The measurement of pulsar J0740+6620 by  Neutron Star Interior Composition Explorer (NICER): 
   $ { M= (2.072 \pm 0.066) M_{\odot}} 
 $   \citet{2021ApJ...918L..27R} and observation of PSR J0030+0451 \citep{2019ApJ...887L..24M}.
\end{itemize}  
\begin{itemize}
    \item The tidal deformabilities observation of neutron star by GW170817 \citep{PhysRevLett.121.161101}.
\end{itemize}
\begin{itemize}
    \item The mass-radii measurement of J1731-347 by HEES \citep{2022NatAs...6.1444D}: $ {M = 0.77^{+0.20}_{-0.17} M_{\odot}}$ and $ {R = 10.4^{+0.86}_{-0.78} }$km.    
\end{itemize}
In Fig. \ref{fig:my_label}, we show the mass-radius (M-R) relation provided by our chosen EOS. HESS J1731-347 could be a quark star, see \citet{2022arXiv221107485D}. Thus, we draw a pure quark M-R relation with bag constant $ {C^{\frac{1}{4}}=150}$MeV. It should be noted that the source PSR J0740+6620 has a 346 Hz spin frequency \citep{2021ApJ...918L..27R}, and the curve shows a static star M-R relation. When spin is taken into consideration, the resulting curve may reach up to 2$ {M_{\odot}}$. In conclusion, the EOSs chosen here are consistent with the observational constraints. \footnote{Notice that the recent observations on neutron stars have ruled out some EOSs, such as the ones used in \citet{2011RAA....11..679X}. They have been successfully used to explain the glitches. We were trying to consider the same scenario to model the energy resource for FRBs and SGRs, by using the same EOSs. It turns out that the scenario is still fine, while the EOSs should be substituted by some other proper ones. }

\section{Application to SGR J1935+2154}

\begin{figure}
    \begin{center}
    \includegraphics[width=0.65\linewidth]{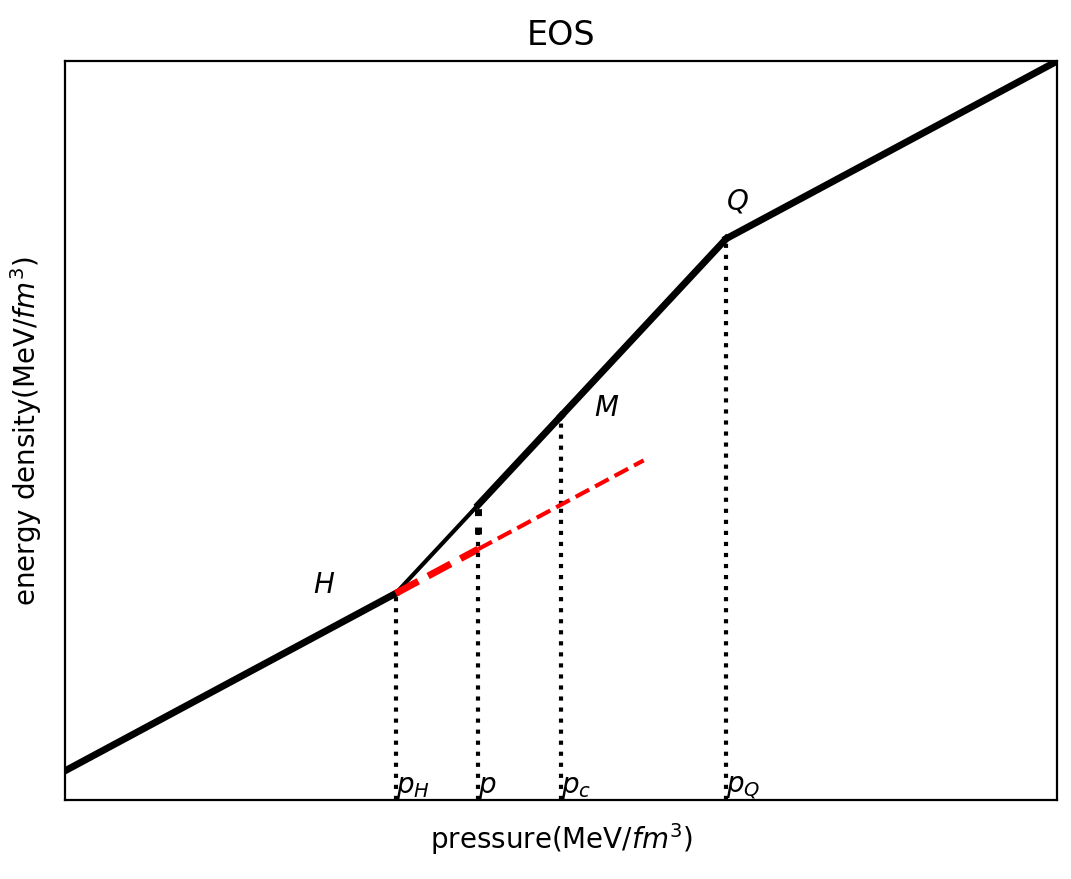}
    \end{center}
    \caption{Schematic plot of EOS for mixed phase, which corresponds to the region of phase transition in Fig. \ref{Fig.sub.2}. The red dashed line (in region of $p_H-p_Q$) 
    corresponds to the metastable state. The black solid lines correspond to the stable state. The dashed vertical line corresponds to the phase transition. In general, the phase transition can happen at any pressure $(p_H<p<p_Q)$, while $p=p_c$ is the typical transition point. The reverse process happens for pressure decreasing, and the quark phase will become metastable state. \textbf{Note that the plot is schematic, and the segment $HQ$ we selected is nearly vertical. The slope value $k$ is approximately $10^9$, which differs from the illustration shown.}
    \label{fig2}}
\end{figure}

SGR J1935+2154, the X-ray, $\gamma$-ray transient sources, is a young NS with a strong magnetic field. It has a spin period $P=3.2$s \citep{2016MNRAS.457.3448I}. Its magnetic field will emit radio pulses and angular momentum will be taken away by open magnetic field lines, so the spin velocity will slowly decrease ($\frac{dP}{dt}\approx 10^{-11}\mathrm{s/s}$) \citep{2016MNRAS.457.3448I}. Because the star's centrifugal force and pressure are both against gravitation, once the star spins down, the centrifugal force will get weaker and the pressure will increase. Consequently, matter in critical pressure will get unstable and transit into the quark phase from the hadronic phase. 
Its stable-state EOS of mixed-phase is shown as the dotted line in Fig. \ref{Fig.sub.2} and also shown as black solid line in Fig. \ref{fig2}. Its critical metastable state of mixed-phase (about to transit its phase) corresponds to a metastable region shown as red dashed line $p_H\rightarrow p_c$ in Fig. \ref{fig2}. Because we do not know the mass of SGR J1935+2154, using the EOS, we choose a boundary condition ($ {p|_{r=0}=70\mathrm{MeV/fm^3}}$) and then integrate the TOV equations.

\subsection{Process of phase transition}
A rotating NS space-time can not be described by Schwarzschild metric, but SGR J1935+2154's angular velocity is 1.93 rad/s only. The slow rotation effects could be a perturbation. \citet{1967ApJ...150.1005H} researched that and calculated the slow spin motion star metric. The metric is given as \citet{1967ApJ...150.1005H} Eq. (66). It is a Schwarzschild metric plus perturbation term that can describe this condition. Using the EOS calculated in Section \ref{2} ($p=f(\rho)$), and the TOV equations, a system of differential equations, the NS's structure can be solved.
\citet{2005PhRvD..72d4028B} had calculated the $\Delta p$ (the perturbative of pressure because of the spin) by the effect of slow rotation and showed the perturbative approach is reliable. 

The picture of how the phase transition occurs is described in the following (similar schematic pictures were given in figs. 2 and 3 of \citet{2011RAA....11..679X}). At the beginning $t=t_i$, the NS EOS is in stable mixed phase, which is shown in Fig. \ref{fig2} as the black solid segments (including  the thick solid segment for stable hadronic part, the thin and thick solid segments for the stable mixed part, and the thick solid segment for the stable quark part). All the hadronic matters are in stable state. With the spinning down of the NS, the pressure rises. Matter in metastable shell boundary (closer to the core) beginning with pressure $p_H$, will extend following the red dashed line (Metastable state, which is discussed in Section \ref{hehehe}). The matter with pressure $p>p_H$ follows the stable state, i.e., the solid line. At a certain time, the whole EOS is corresponding to the thick segments (including the thick solid segment for stable hadronic part, the thick dashed segment for the metastable mixed part, the thick solid segment for the stable mixed part, and the  thick solid segment for the stable quark part). If the rising of pressure is not significant, these matters will still be in a hadronic matter state metastably (red dashed line). With the increase of pressure, the metastable matter tends to the transition. At time $t=t_{i+1}$, metastable shell boundary pressure $p_H$ increases to a certain trigger pressure $p$, which is random in the range of $(p_H, p_Q)$, while $p_c$ could be the typical value. All the metastable matter transits to the stable state, i.e., the thin solid line in Fig. \ref{fig2}. 
The transition matter density increases, and the volume decreases. The core will shrink, and consequently the gravitational potential energy will be released. This energy could be the source of the FRBs and the SGRs. 
The process repeats to the next circle after the phase transition.

Now we calculate the potential energy release from the phase transition. To begin with the unstable critical state, metastable state EOS is shown in Fig. \ref{fig2} the red thick dashed lines, and other regions are stable, which are shown in Fig. \ref{fig2} as the solid lines. Solving TOV equations with this EOS and boundary condition $p|_{r=0}=70 \mathrm{MeV/fm^3}$, we can get the structure of the critical metastable NS. Then, the position of $p_H$ is known and denoted by $r_i$. The star spins down slowly, and it decreases by $\frac{dP}{dt} t$. We simplify the spin-down process that it only increases the pressure in different radii to balance the gravity. After passing time $t_p$ ($t_p=t_{i+1}-t_i$), $p(r_i)$ increases to $p_c$, the metastable matter will transit its phase into the stable EOS and the next cycle begins. For each phase transition, the released gravitational energy transfers to other radiations like FRBs and SGRs gradually, and the magnetar becomes active. $t_p$ can be taken as the time interval between two active windows. 

Phase transition transforms the hadronic phase matter into quark matter. These two-phase densities are different at the same pressure. The change of neutron star matter density deduces the change in volume. Assuming $r_m$ is the thickness of metastable matter, the magnitude of the $r_m$ shrink during the phase transition is: 
\begin{equation}
    \Delta r=r_m-\frac{\rho r_m}{\rho'},
\end{equation}
where $\rho$ is the average density of metastable matter, and $\rho'$ is the average density of stable matter 
which forms after phase transition at $r_m$. After knowing the shrink, we can calculate the gravitational potential energy released. 
To calculate the released energy, a precise calculation involves determining the structures of both metastable and stable states of the whole star. And then calculate the difference of the gravitational potential energy of these two states. However, the gravitational potential energy is in the order of $ {10^{54}}$ erg, while the released gravitational energy is as low as $ {10^{43}}$ erg (see Table \ref{tab}). As a result, the numerical error is enormous. 
Therefore, we have to make some simplifications to calculate the released energy.

We numerically divide the star into many layers. To make the calculation affordable, the thickness of each layer is in order of 1 m. It is thicker than the the thickness of metastable matter $r_m$. Then, we calculate the shrink of every layer caused by particle drop towards the NS core. Particles replace process obeys $ {r^2 n(r) \delta r }= $constant, where the $ {n(r)}$ is the particle number density at radius $r$, and $ {\delta r} $ is the drop distance of each layer. This equation keeps the particle number conversation during the shrink.
At this stage, we assume the $ {n(r)}$ is the same as in metastable state.
We calculate the gravitational potential energy released in every position $r\rightarrow r+dr$, and sum them up. This is the total energy released in each phase transition.
For the next phase transition, we should recalculate the whole structure and also the $n(r)$ by solving the TOV equations, and then repeat the whole process.

We choose different values of the free parameter $k$ and the outcome is shown in Table~\ref{tab}. For a given $k$, all other four quantities are determined. We choose the parameter $k$ varies a good variety, and mainly concentrate at around $10^9$, where the outcome is comparable to the observations of SGR J1935+2154, i.e., $t_p \sim 1$ year, released energy is in order of $ {10^{43}}$ erg. The $\Delta r$ is about $ {10^{-7}}$m when $k$ is about $ 10^9$. Compared with NS radius, $r_m$ is tiny. Thus, the phase transition is repeatable. All these values are in a reasonable region. Notice that all the calculations only rely one free parameter, i.e., $k$.
We also calculated some other EOSs models for hadronic matter (see Table~\ref{tab2}). It shows all of the three models can produce required energy and recurrence time interval. This indicates the phase transition scenario for the FRB and SGR energy source is not highly model dependent.

\begin{table}
\begin{center}
\begin{tabular}{ccccccc}
   \toprule
  $k$  & $\epsilon$ & $p_c - p_H$ ($\mathrm{MeV/fm^{3}}$) & $t_p$ (year) & energy (erg) & glitch  \\
   \midrule
  
$2\times 10^9$ & $\epsilon_1$ & $6.5\times 10^{-9}$&  $0.23$ & $9.0\times 10^{42}$ &  $-$ \\

   $ 2\times 10^9$ & $\epsilon_2$ & $1.3\times10^{-8}$&  $0.45$ & $3.6\times 10^{43}$ &  $-$ \\

   $ 2\times 10^9$ & $\epsilon_3$ & $2.0\times10^{-8}$&  $0.70$ & $8.1\times 10^{43}$ &  $-$ \\

$2\times 10^6$ & $\epsilon_1$ & $6.5\times 10^{-6}$&  $\sim 10^3$ & $9.0\times 10^{45}$ &  $\sim 10^{-7}$ \\

   $ 2\times 10^6$ & $\epsilon_2$ & $1.3\times10^{-5}$&  $\sim 10^3$ & $3.6\times 10^{46}$ &  $\sim 10^{-7}$ \\

   $ 2\times 10^6$ & $\epsilon_3$ & $2.0\times10^{-5}$&  $\sim 10^3$ & $8.1\times 10^{46}$ &  $\sim 10^{-7}$ \\

   $2\times 10^3$ & $\epsilon_1$ & $6.4\times 10^{-3}$&  $>10^4$ & $1.0\times 10^{49}$ &  $\sim 10^{-5}$ \\
   $ 2\times 10^3$ & $\epsilon_2$ & $0.013$&  $>10^4$ & $4.0\times 10^{49}$ &  $\sim 10^{-5}$ \\
   $ 2\times 10^3$ & $\epsilon_3$ & $0.02$&  $>10^4$ & $8.9\times 10^{49}$ &  $\sim 10^{-5}$ \\
   \bottomrule
\end{tabular}
\caption{The outcome of phase transition calculation with different value of free parameter $k$. $p_c$ is the cross point in Fig. \ref{Fig.sub.1} (\textbf{EOS DD2} ). $\epsilon_1 $ denotes that the energy density of cross point chose as $[ 3\epsilon_{\mathrm{H}}(p_c) + \epsilon_{\mathrm{Q}}(p_c) ] / 4,$ $ {\epsilon_2}$ denotes that the energy density of cross point chose as $ {[\epsilon_{\mathrm{H}}(p_c)+\epsilon_{\mathrm{Q}}(p_c)]/2}$, and $ {\epsilon_3} $ denotes that the energy density of cross point chose as $ {[\epsilon_{\mathrm{H}}(p_c)+3\epsilon_{\mathrm{Q}}(p_c)]/4}$.  With a given $k$, one can get $p_H$ from the cross point between the dotted line and the hadronic line in Fig. \ref{Fig.sub.2}. $t_p$ is the period between the two phases' transition. Energy is the released gravitational potential energy. The approximate magnitude of the glitch is the glitch during the phase transition. These three quantities are all able to calculate based on the method described in the text, and they can be directly  compared with the observations. `` -'' denotes that numerical error enormous, and the result is inaccurate.} 
  \label{tab}
\end{center}
\end{table}

\begin{table}
\begin{center}
\begin{tabular}{ccccccc}
   \toprule
EOS name & $t_p$ (year) & energy (erg)  & k & $ {p|_{r=0}\mathbf{(MeV/fm^3)}}$ \\
   \midrule
  
DD2 & $0.45$ &  $3.6\times 10^{43}$  & $2\times 10^9$   &    \textbf{70}\\  
DS (CMF)-5 & $ \boldsymbol{1}$ &  $ \boldsymbol{6.4\times 10^{43}}$  & $ \boldsymbol{2\times 10^8}$  & \textbf{70}\\
JJ(VQCD) & $ \boldsymbol{0.67}$ &  $ \boldsymbol{4.9\times 10^{43}}$ & $ \boldsymbol{2\times 10^9}$ & \textbf{100} \\

   \bottomrule
\end{tabular}
\caption{Some hybrid star model EOSs of hadronic. This calculation outcome shown here is to make sure our model not highly model-dependent.}
  \label{tab2}
\end{center}
\end{table}

\subsection{SGR J1935+2154}
We apply the model to the magnetar SGR J1935+2154. This $t_p$ is calculated and should be consistent with the observation. Its continual spin-down may induce inner core matter phase transition, and we calculate how much gravitational potential energy can be released. This energy is taken as the main energy source for one active window released in this process. 
The property of metastable shell is related to surface tension $\sigma$ and $k$ and we choose a typical critical pressure $p= {\boldsymbol{48.53}}  \mathrm{MeV/fm^{3}}$. Once the metastable state reaches this pressure, the transition occurs. For SGR J1935+2154, the $\frac{dP}{dt}=1.43 \times 10^{-11} \mathrm{s/s}$, $B=2.3\times 10^{14}$ Gauss, $P=3.2 \mathrm{s}$, where $B$ is the magnetic field and $P$ is the period of spin \citep{2016MNRAS.457.3448I}. 

As discussed above, we use the critical metastable states as the initial condition and get an initial structure. 
The EOS of metastable state in this region is hadronic as shown in Fig \ref{fig2}, donated as the red dashed line. 
After the phase transition, the star will be in a stable state, and the EOS and structure of the NS get changed. Using the two different EOSs and TOV equations, the NS's structure of metastable state and the structure of stable one can be calculated. Thus, we get the magnitude of the NS core shrink. Finally, we can calculate the released gravitational potential energy. 

The total energy of an SGRs is about $4.8 \times 10^{40}$ erg \citep{2020ApJ...904L..21Y}. The energy of an FRBs is $3 \times 10^{34}$ erg \citep{2020Natur.587...59B}. As shown in Table \ref{tab}, with proper choosing of free parameter $k$, the released energy from gravitational shrinking is in order of $10^{44}$ erg. This energy is enough for powering all the FRBs and SGRs during an active interval.
Phase transition latent heat is about $\Delta \mu r_c^2  n  {r_m} \sim 10^{34}$erg (for $k = 2 \times 10^{10}$), which is tiny comparing with the potential energy. It is neglected in this work. The change of rotational kinetic energy is $J \Delta \Omega \sim 10^{30}$ erg (for $k=2 \times 10^{10}$). It is also ignored.

 The stable SGR J1935+2154 slowly spins down. 
When the star reaches the critical metastable state, the interval between two states is, $t_p$, which is the interval between two active windows. The observation of SGR active window's period is about 238 days \citep{2021ApJ...923L..30Z}. This corresponds to $k=2 \times 10^9$, where the $t_p\sim0.45$ year, which is roughly consistent with the active window. If we choose the phase transition at $p=p_c$, the thickness of the transition shell $r_m \sim  { 10^{-6}}$ m, which means the number of phase transitions is far enough for the repeating FRBs and SGRs.

By comparing two states before and after the transition moment of inertia, one can also approximately calculate the magnitude of the glitch. 
We choose three components to compare with observations, $t_p$, energy, and glitch. We find that with a large $k$ (about $10^9$), energy is enough for one SGR active window. $t_p$ is consistent with the observation. The magnitude of the glitch is appropriate. With a small k, the energy is enough, but $t_p$ is too long, and the glitch is huge.

\section{Conclusion and Discussion}
In this work, we choose the DD2 model and MIT bag model to describe the hadronic and quark matter and numerically calculate the EOS of hadrons and quarks. Then, we get the structure of the stable and metastable NS with mass of 1.62$M_{\bigodot}$. With the spin-down effect, hadrons reach the critical pressure and transit their phase. We compare the theoretical calculation with the observations of SGR J1935+2154 and find that if a large $k$ is chosen, the plausible phase transition model can produce enough energy to power the SGRs and the FRBs. The outcome is consistent with observations including the total energy, repeating period, and also the glitches. We suggest that the phase transition could be a plausible model for both SGRs, as well as the cosmological repeating FRBs, where their SGRs are too weak to be observed. We also test different EOS models, and find it is not highly model dependent.

In this article, we only utilized one free parameter $k$, and the outcome is well consistent with the observations. Another uncertainty comes from the boundary condition. We choose $p|_{r=0} = 70 \mathrm{MeV/fm^3}$. This condition induces the NS mass being 1.62 $M_{\bigodot}$. If the mass can be determined by other method, the boundary condition become determined. 

This phase transition also can be the source of FRB, but here we have not studied the radiation mechanism yet. It could be similar to the model described in \citet{2020MNRAS.498.1397L}, which discussed how the energy transforms to radio pluses and X-ray emission. The disturbance will conduct to charge starvation region and their FRBs will be produced by coherent emission. People are searching gravitational wave transients associated with SGR J1935+2154 active \citep{2023IAUS..363..187M}. We can calculate the GW from the structure changed by phase transition, and compare it to observations of gravitational wave.

Very recently, SGR J1935+2154 has been reported two giant glitches with magnitude $\frac{\Delta \nu}{\nu}$ being $\sim 3.1 \times 10^{-5}$ and $\sim 6.4 \times 10^{-5}$, respectively  \citep{2022arXiv221103246G}, and a giant anti-glitch with magnitude $\frac{\Delta \nu}{\nu}$ being $\sim 5.8 \times 10^{-6}$ \citep{2022arXiv221011518Y, 2022arXiv221108151W}. Similar to the giant glitches in radio pulsars, they are likely originated from super-fluid mechanism \citep{1996ApJ...459..706A, 2015IJMPD..2430008H, 2021MNRAS.507.2208W}. The spin-down slopes are recovered after those giant glitches \citep[see][for example]{1990Natur.345..416F}. These giant glitches do not release a noticeable amount of energy. The structure as well as the pressure and density profile may not change much before and after the giant glitches, which means they do not affect the normal spin-down-induced phase transition glitches. In other words, the giant glitches do not affect the overall spin-down trending, and therefore, the spin-down induced small glitches keep the same amount of released energy and the same intervals as given in Table \ref{tab}. The anti-glitches slow down the NS at a faster rate, which might lead to a quicker phase transition. The active period (i.e., SGR outbursts as well as FRBs) may start earlier and the intervals between successive active periods become shorter. As the anti-glitch is unpredictable, it may increase the randomness of the intervals.

\section*{Acknowledgements}
We are very grateful for the anonymous referee, especially for suggesting to check the validity of the EOSs we used. It turned out the former EOS based on the mean field approximation model was not able to fulfil the new observational results as shown in Figure \ref{fig:my_label}. We thank the helpful discussions with Weihua Lei, Qingwen Wu, Shiyan Tian, Yunwei Yu, Jumpei Takata and Enping Zhou.
This work is in part supported by the National Natural Science Foundation of China (Grant Nos. 12041306 and U1931203), and by the National Key R\&D Program of China (2022SKA0130103 and 2021YFA0718504). We also acknowledge the science research grants from the China Manned Space Project with No. CMS-CSST-2021-B11.



\bibliography{frb}{}
\bibliographystyle{aasjournal}
\end{document}